\documentclass[sigconf]{acmart}

\usepackage[textsize=small]{todonotes}
\usepackage{amsmath}
\usepackage[mode=buildnew]{standalone}
\usepackage{multirow}
\usepackage{color}
\usepackage[normalem]{ulem}
\usepackage{adjustbox}
\usepackage{xspace}

\newcommand{\mb}{\mathbf}
\newcommand{\mcal}[1]{\mathcal{#1}}

\newcommand{\qenc}{E_Q}
\newcommand{\penc}{E_P}
\newcommand{\h}[1]{\textbf{#1}}

\newcommand{\concurrentqa}{\textsc{ConcurrentQA}\xspace}

\usepackage{enumitem}

\AtBeginDocument{%
  }

\setcopyright{acmlicensed}
\acmConference[REML '23]{the 2023 ACM SIGIR Workshop on Retrieval-Enhanced Machine Learning}{July 27, 2023}{Taipei, Taiwan}
\acmYear{2023}
\copyrightyear{2023}
\acmDOI{}
\acmISBN{}
\acmPrice{}

\begin{document}

\title{Resources and Evaluations for Multi-Distribution Dense Information Retrieval}

\author{Soumya Chatterjee}
\affiliation{%
  \institution{Stanford University}
  \city{Stanford}
  \state{California}
  \country{USA}
}
\email{soumyac@stanford.edu}

\author{Omar Khattab}
\affiliation{%
  \institution{Stanford University}
  \city{Stanford}
  \state{California}
  \country{USA}
}
\email{okhattab@stanford.edu}

\author{Simran Arora}
\affiliation{%
  \institution{Stanford University}
  \city{Stanford}
  \state{California}
  \country{USA}
}
\email{simarora@stanford.edu}

\renewcommand{\shortauthors}{Chatterjee et al.}

\begin{abstract}
We introduce and define the novel problem of multi-distribution information retrieval (IR) where given a query, systems need to retrieve passages from within multiple collections, each drawn from a different distribution. Some of these collections and distributions might not be available at training time. To evaluate methods for multi-distribution retrieval, we design three benchmarks for this task from existing single-distribution datasets, namely, a dataset based on question answering and two based on entity matching. We propose simple methods for this task which allocate the fixed retrieval budget (top-$k$ passages) strategically across domains to prevent the known domains from consuming most of the budget. We show that our methods lead to an average of 3.8+ and up to 8.0 points improvements in Recall@100 across the datasets and that improvements are consistent when fine-tuning different base retrieval models. Our benchmarks are made publicly available.

\end{abstract}


\maketitle

\begin{figure*}[tbp]
    \centering
    \includegraphics[width=0.9\textwidth]{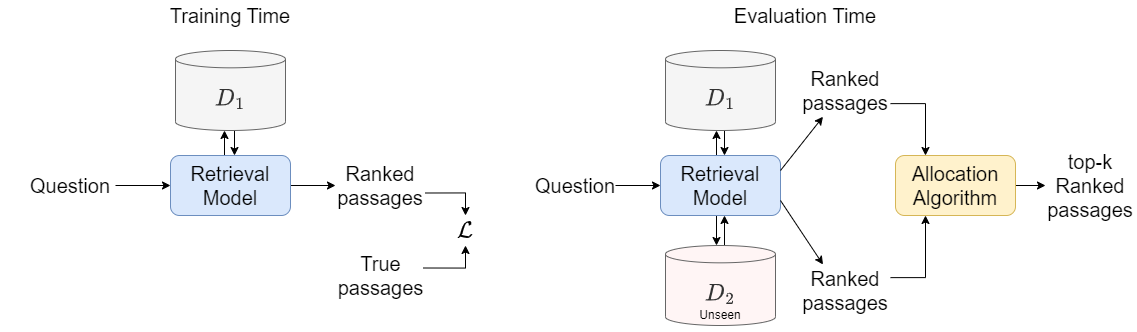}
    \caption{Overview of the multi-distribution retrieval task. During training, some distributions $\mcal{D}_1$ are available. During evaluation, queries which require retrieval from the seen distributions $\mcal{D}_1$ and unseen distributions $\mcal{D}_2$ are asked. Different allocation algorithms are used to retrieve a fixed number $k$ of passages by combining the retrieved passages from $\mcal{D}_1$ and $\mcal{D}_2$. This is needed so that passages from $\mcal{D}_1$, which the model was trained on, do not consume most of the budget $k$.}
    \label{fig:overview}
\end{figure*}

\section{Introduction}

A key feature of retrieval systems is the ability to incorporate fresh data in non-stationary domains such as news and social media \citep{bendersky2023reml}. Over the lifetime of a system, users may also want to incorporate entirely new types of information such as new code libraries developed over time \citep{zan2022language}. The challenges of reasoning over non-stationary information are well-studied when the information is being encoded within the parameters of a deep-learning model. For instance, challenges include catastrophic forgetting and expensive retraining \citep{hoffart2014discovering, hoffart2016knowledge, jang2021towards, gadde2021towards}. However, the properties of non-stationary retrieval are comparatively underexplored. In a real-world retrieval system, the retrieval model might view \textit{some, but not all} of the inference time data distributions during training time. We study on this underexplored \textit{multi-distribution} retrieval setting.

In this work, we suggest that the standard dense retrieval protocol may not always be preferable in the multi-distribution retrieval setting. Concretely, the standard retrieval protocol with dense retrievers entails: (1) encoding the user question and passages in the corpus $\mathcal{C}$ with the dense model $\mathcal{M}$, (2) computing similarity scores between the question embedding and passages in $\mathcal{C}$, and (3) returning the passages with the top-$k$ highest similarity scores to the user \citep[inter alia.]{chen2017retrieveread, karpukhin2020dpr, lewis-etal-2021-paq}. Suppose $\mathcal{C}$ consists of many distributions of passages $\{D_1, D_2, D_3\}$ (e.g. Wikipedia, StackOverflow, and Ubuntu logs) and the retriever $\mathcal{M}$ is trained on documents that are similar to a subset of these distributions $\{D_1\}$ (e.g. Wikipedia). We ask whether $\mathcal{M}$ is \textit{biased} towards retrieving from $\{D_1\}$ such that naïvely selecting the documents with the top-$k$ documents will lead to over-retrieving from $\{D_1\}$ and under-retrieving from $\{D_2, D_3\}$? 

To identify whether the retriever is biased towards a particular distribution, we propose to evaluate retrievers on questions that rely retrieving an equal number of passages from each distribution in order to answer the question. For instance, consider the question ``Which store offers a better deal on Apple products?'', which requires that the retriever retrieve passages about Apple products from multiple online retailers, each of which uses a different style for their listing.
Unfortunately existing retrieval benchmarks are not designed to evaluate retrievers in this multi-distribution setup. Existing benchmarks generally require the retriever to operate over a corpus $C$ containing passages from a single static distribution of text $C = \{D_i\}$ \citep{thakur2021beir}. While existing benchmarks facilitate training the retriever on questions over $C = \{D_i\}$ and evaluating with questions over $C = \{D_j\}$, the prior benchmarks do not contain questions that simultaneously require a passage from $\{D_i\}$ and $\{D_j\}$ to produce an answer.

To facilitate evaluating retrievers on corpora consisting of multiple passage distributions, we contribute a suite of three new benchmarks ($D_i-D_j$): Walmart-Amazon, Amazon-Google and Wikipedia-Enron. The benchmarks are constructed by modifying existing entity matching \citep{magellandata} and multi-hop question answering \citep{arora2022reasoning} datasets. In our benchmarks, each question requires retrieving one passage from $D_i$ and one from $D_j$ in order to answer the question. We consider training retrievers on $D_i$ and evaluating on the mixture of $\{D_i, D_j\}$ at inference time.
Using this protocol, we validate our hypothesis that naïvely retrieving the  passages with the top-$k$ highest similarity scores over-retrieves from $D_i$ and under-retrieves from $D_i$. Concretely, we find that instead allocating the fixed retrieval budget (i.e. $k$ passages from $\{D_i, D_j\}$) to distributions (i.e. enforcing $k_i$ passages from $D_i$ and $k_j$ passages from $D_j$) leads to a 3.8+ Recall@10 (5.7\%) improvement on average.

\paragraph{Contributions}
\begin{itemize}
    \item We create three benchmarks for the underexplored task of multi-distribution retrieval where certain inference time distributions are unseen at train time. Our benchmarks build on existing entity-matching and question answering datasets.
    \item We experiment with simple methods for multi-distribution retrieval that strategically allocate the fixed retrieval budget across distributions and report performance on the three benchmarks.
    \item We perform thorough analysis and ablation studies to gain insight into the benchmarks and allocation strategies. Concretely, we investigate the effect of size of retrieval budget, size of training set, and choice of base pretrained models.
\end{itemize}

We hope the resources and evaluations we provide facilitate future research on retrieval-enhanced machine learning systems in non-stationary domains.\footnote{Our data is available at \url{https://github.com/stanfordnlp/mixed-distribution-retrieval}.}

\begin{table*}[htpb]
    \centering
    \resizebox{0.85\textwidth}{!}{%
    \begin{tabular}{|l|l|}
    \hline
    Query    &  Do ShopBack and Cognitive Arts both deal with internet based services?\\
    Passage1 (Wiki) & ShopBack is a Singaporean-headed e-commerce startup that utilises the cashback reward program. \\
    & It allows online shoppers to take a portion of their cash back when they buy products through \textellipsis \\
    Passage2 (Email) & Cognitive Arts, a developer of Internet-based products and services for educational and corporate \\
    &training uses, said it appointed Russell C. White as chief executive officer \textellipsis \\
    \hline
    Query    &  How do the Walmart and Amazon listings of `Acer Iconia Tablet Bluetooth Keyboard' differ?\\
    Passage1 (Walmart) & Acer Iconia Tab Bluetooth Keyboard Bluetooth model 2.0 Removable AAA Battery is included \\
    & 32.8 operating distance LED power pairing battery indicator Thin stylish design Convenient \textellipsis \\
    Passage2 (Amazon) & The Bluetooth Keyboard for the Acer Iconia Tab is the perfect accessory for increased productivity. \\
    & Wirelessly connect to your Tab for seamless typing and navigation. This slim keyboard is the \\
    & perfect travel companion for when you take your Tab on the road. It conveniently fits in \textellipsis \\
    \hline
    Query    &  Are Eros International and MicroEmissive Displays in the same type of industry? \\
    Passage1 (Wiki) & Eros International PLC is a leading global company in the Indian film entertainment industry, \\
    & the Isle of Man. Through its production and distribution subsidiary, Eros International, it \textellipsis \\
    Passage2 (Email) & MicroEmissive Displays, which develops microdisplays for embedding into portable electronics\\
    & products, said it raised GBP 1 million (\$2.1 million) in its first round of funding \textellipsis \\
    \hline
    Query    &  Which of American Fur Company or Hyperchip Inc. was founded first?\\
    Passage1 (Wiki) & The American Fur Company (AFC) was founded in 1808, by John Jacob Astor, a German \textellipsis\\
    Passage2 (Email) & Richard Norman, president and CTO of Montreal-based Hyperchip Inc \textellipsis\ since co-founding \\
    & the company back in 1997, he has averaged about 100 hours a week \textellipsis\\
    \hline
    \end{tabular}
    }
    \caption{Examples of queries and relevant passages from our datasets.}
    \label{tab:dataset-examples}
\end{table*}

\section{Related Work}

\paragraph{Retrieval-Based Systems} Several NLP applications including open-domain question answering (QA) \citep{voorhees1999trec} and personal assistants \citep{nehring-etal-2021-combining} require the ability to handle a wide range of topics. This capability is usually added in one of two ways -- implicit-memory methods which make model parameters `remember' information, and retrieval-based methods which learn to fetch relevant documents from a corpus (like Wikipedia) or a knowledge graph. We focus on retrieval-based methods, which have been shown to outperform current implicit methods on open-domain applications \citep{lewis-etal-2021-paq, borgeaud2021retro}. Retrieval-based systems typically consist of a retriever and a reader \citep{chen-etal-2017-reading}. Under standard retrieval mechanics, the retriever performs maximum inner product search between the question embedding and all passages embeddings and returns the passages with the top-$k$ inner product scores. 

In general, existing  benchmarks require retrieving passages from a corpus with a single distribution of passages \citep{thakur2021beir, yang2018hotpotqa, rajpurkar-etal-2016-squad, bajaj2016ms}. Related to our task, \citet{arora2022reasoning} created a dataset \concurrentqa\ for retrieval from multiple (public and private) distributions. The proposed benchmark contains multi-hop questions requiring arbitrary retrieval over the distributions. To study the retrievers' bias to each sub-distribution in a principled manner, we isolate questions that require one passage from each sub-distribution in our modified benchmark. We also evaluate a broader set of models and ablation settings. Related to our task, \citet{chen2020hybridqa} created a dataset called HybridQA for retrieval over tables and text. We focus on evaluating over multiple textual distributions, as many off-the-shelf retrievers are trained over text \citep{karpukhin-etal-2020-dense, reimers-2019-sentence-bert}. Finally, prior work proposes benchmarks for non-stationary QA in which values within pages change over time \citep{zhang2021situatedqa, wang2021temporalqa}. For instance, the ``location of the next Olympic games'' changes over time. Our benchmarks explore a different type of dataset shift where the format of passages and/or overall set of topics being discussed differ across distributions. 

\paragraph{Domain Adaptation} Our problem setting bears resemblance to those of domain adaptation \citep{ben2010theory} and out-of-domain retrieval generalization \citep{thakur2021beir}. In domain adaptation, a model trained on one domain or distribution is adapted to work well on another with the goal of transferring knowledge from a source domain with sufficient labelled data to an unlabelled target domain. This is challenging since neural networks are sensitive to distribution shifts. Several methods like learning domain independent features \citep{JMLR:v17:15-239} or adapting based on unlabelled target domain examples \citep{liu2019transferable} have been proposed. However, in our setting, no examples from the unseen distributions are available and unlike domain adaptation, we want our models to perform well on a spectrum of distributions that are similar-to-dissimilar to the training distribution, rather than only performing well on the target distribution.

\section{Problem Definition}
In this section we define the task of Multi-Distribution Retrieval and describe the benchmarks we propose for it.

\subsection{Setup}
\label{sec:task-defn}
We consider the multi-distribution retrieval problem, where the retrieval corpora comes from different distributions, only a subset of which is available during model training. For example, we might have two data distributions $\mcal{D}_1$ and $\mcal{D}_2$ where $\mcal{D}_1$ is known during training while $\mcal{D}_2$ is not (Figure~\ref{fig:overview}). Possible reasons for $\mcal{D}_2$ not being available include it being a private dataset on which we cannot train or it being produced over time or in the future (i.e., after the model has been trained).

The two distributions are two sources of text passages with different characteristics. For example, one distribution could be Wikipedia passages and the other could be email snippets. The passages from these two sources can be expected to form two different distributions since the encyclopediac nature of Wikipedia passages would be different from the conversational nature of emails. In this example, it could be that such emails are never available during training due to privacy reasons.

Formally,
we have sets of text passages $D_1 = \{d^{1}_{1}, d^{1}_{2}, \ldots d^{1}_m\}$ and $D_2 = \{d^{2}_{1}, d^{2}_{2}, \ldots d^{2}_n\}$ with different characteristics (e.g. Wikipedia passages and email snippets) drawn from $\mcal{D}_1$ and $\mcal{D}_2$ respectively. Here only $D_1$ is available during training while $D_2$ is not.
Given a query $q$, the multi-distribution retrieval task requires retrieving two passages $d^1_i$ and $d^2_j$ from the two corpora $D_1$ and $D_2$ respectively which are most relevant to the given query $q$. Examples of queries and relevant passages are shown in Table~\ref{tab:dataset-examples}.

\subsection{Benchmarks}
One of our primary contributions is the creation of benchmarks for our novel multi-distribution retrieval task owing to the lack of existing ones. We adapted datasets from prior work on different tasks like question answering and entity matching in order to create our benchmarks. Particularly, we use a modified version of \concurrentqa~\citep{arora2022reasoning} which is a dataset constructed for investigating privacy-preserving QA. It consists of multi-hop questions over Wikipedia and Enron emails corpora (forming two distinct distributions). We use a subset of questions (called comparison questions\footnote{e.g., ``Do ShopBack and Cognitive Arts both deal with internet based services?''}) which can be answered in a single-hop and which require passages to be retrieved from both the corpora. There are 100 such question in the validation and test sets each. The Wikipedia and emails corpora have 5.2M and 47k passages respectively. There are roughly 4000 question which require retrieval from one corpus only and we use these for training our models.

The second dataset is created using entity-matching datasets from \citet{magellandata}. We use the Walmart--Amazon dataset where given a product title, the goal is to retrieve its description from both sources. We provide the scatter plot of BERT embeddings of the product descriptions in Figure~\ref{fig:bert-tsne} to show that they represent different distributions. The Amazon and Walmart corpora have 21891 and 2520 products respectively. We also have a mapping between the listing of the same item in the two sites for 1127 items which we use for evaluation (split between validation and test sets in a 1:1 ratio). There are actually 2$\times$1227 queries since product title of either source can be used as the query. The unmatched products are used for training the retrievers. When Amazon is assumed to be the known distribution, we train on (title, description) tuples from Amazon and evaluate on the matched items.

Finally, the third dataset is based on the Amazon--GoogleProducts entity-matching dataset from \citet{kopcke2010evaluation} which is similar to the Walmart--Amazon dataset above. The Amazon and Google corpora have 1248 and 3035 products respectively and a mapping between 1161 products is also present. All dataset statistics are summarized in Table~\ref{tab:datasets}.

\begin{table}[htbp]
    \centering
    \resizebox{\linewidth}{!}{%
    \begin{tabular}{|l|cccc|}
        \hline
        Dataset & $|D_1|$ & $|D_2|$ & Num val & Num test \\ \hline
        Walmart--Amazon & 2520 & 21891 & 1127 & 1127 \\
        Amazon--Google & 1248 & 3035 & 1161 & 1161 \\
        \concurrentqa & 5.2M & 47k & 100 & 100 \\
        \hline
    \end{tabular}
    }
    \caption{Number of passages in each corpus and the number of queries for the three datasets used.}
    \label{tab:datasets}
\end{table}

\section{Methods}
In this section, we give an overview of dense retrieval methods, and describe our training and evaluation protocol. Finally, we discuss our method for multi-distribution retrieval along with baselines.

\subsection{Dense Retrieval Methods}
\label{sec:dense-retrieval-methods}
For dense retrieval from a corpus $\mcal{C}$ based on a query $q$, prior work (e.g. \citealp{chen-etal-2017-reading}, \citealp{karpukhin-etal-2020-dense}) use encoders $\qenc$ and $\penc$ to get the embeddings of the query ($\mb{e}_q$) and all the passages $\left\{\mb{e}_c\ |\ c \in \mcal{C}\right\}$ respectively. The similarity between the query embedding and each of the passage embeddings is computed and the passages with the highest similarity is returned as an answer to the query $q$. The similarity here can be dot product or cosine similarity. Further, typically, top-$k$ passages (for some $k$) are returned instead of one. The query and passage encoders $\qenc$ and $\penc$ usually have the same architecture but may or may not have the same parameters \citep{karpukhin-etal-2020-dense}. It is common to use pretrained Transformer encoders \citep{devlin-etal-2019-bert, liu2019roberta} and fine-tune them for retrieval. The retrieved passages are typically fed to a reader model to generate the answer \citep{chen-etal-2017-reading} but in this work, we focus on only the retrieval part of the process.

For our query and passage encoders ($\qenc$ and $\penc$), we fine-tune RoBERTa encoders \citep{liu2019roberta} with parameters shared between $\qenc$ and $\penc$. Our retrievers are based on those of \citet{xiong2020answering} but modified to our single-hop multi-distribution retrieval task.

\paragraph{Training:}
As discussed in Section~\ref{sec:task-defn}, during training, we have access to data from distribution $\mcal{D}_1$ only in the form of the corpus $D_1$. For this distribution, we also have access to a dataset $Q_1 = \{(q_1, p_{11}, p_{12}, \ldots), (q_2, p_{21}, p_{22}, \ldots), \ldots \}$ of queries $q_i$ and corresponding relevant passages $\{p_{i1}, p_{i2}, \ldots\}$. Such datasets are commonly used in question answering and are readily available.
We use this dataset for fine-tuning our models. For each example in the dataset, we have a query $q_i$. We sample one positive passages $p_{ij}$ from the positive passages corresponding to $q_i$ and also randomly sample another passage $p^{\prime}_{ij}$ from the corpus $D_1$ to act as a negative passage. Using these, the loss on a single example is given by:
\begin{equation*}
    \mcal{L}_{ij} = -\log \frac{e^{\qenc(q_i)^{\top} \penc(p_{ij})}}{e^{\qenc(q_i)^{\top} \penc(p_{ij})} + e^{\qenc(q_i)^{\top} \penc(p^{\prime}_{ij})}}
\end{equation*}
The average of $\mcal{L}_{ij}$ over a batch of $B$ samples is minimized in one training iteration. Other more sophisticated choices of negative examples like sampling hard-negatives using BM25 can also be employed instead for randomly sampling $p^{\prime}_{ij}$.

\paragraph{Evaluation:}
During evaluation, we are given queries $q$ which need to retrieve two passages $p_1$ and $p_2$ from the two corpora $D_1$ and $D_2$ representing the two distributions. Also, recall that we return the top-$k$ passages based on the similarity of their embeddings to the query embedding. Now, since the encoders were trained on examples from $\mcal{D}_1$, we expect it to be proficient in retrieving examples from $D_1$. That is, it will assign high scores to relevant passages from $D_1$ and low scores to the irrelevant ones. However, the encoders, which had not seen passages from $\mcal{D}_2$ during training, will not be so good in retrieving from $D_2$. The scores assigned to the relevant passages might not be as high as scores assigned to some irrelevant passages from $D_2$. 

A simple approach to multi-distribution retrieval would combine the two corpora $D_1$ and $D_2$ to a single corpus $D_{\text{merged}}$ and retrieve from it. However, based on the above observations, we can see that this na\"ive approach will not work since the passages from $D_1$ will have higher scores and use up most of the budget of $k$ passages. A pragmatic approach here would try to balance between the two corpora and allocate portions of the retrieval budget to both $D_1$ and $D_2$. 
To study this hypothesis, we next propose various strategies of allocating the fixed retrieval budget across passage distributions.

\subsection{Allocation Strategies}
Given a retrieval budget of $k$ passages, several allocation strategies described below can be used.

\begin{itemize}[leftmargin=1em, itemsep=0em, topsep=0em]
\item Na\"ive merging: Merge the two corpora $D_1$ and $D_2$ into a single corpus $D_{\text{merged}}$ and retrieve the top-$k$ passages from it. This is the simplest approach and is equivalent to (incorrectly) assuming that the passages $D_2$ from the unseen distribution are also drawn from the same distribution as $D_1$.
\item Per-task allocation: Retrieve $k_1$ passages from $D_1$ and $k_2$ passages from $D_2$ such that $k_1 + k_2 = k$. Here $k_1$ and $k_2$ are the same across different queries. This approach takes into account the fact that $D_1$ and $D_2$ are drawn from different distributions and the model being better calibrated on one, handles retrieval from $D_1$ and $D_2$ differently. We parameterize this method by the fraction $k_1/k$ allocated to $D_1$ in the following sections.
\item Per-query allocation: Per-task allocation done at the level of each individual query. The budget $k$ is divided in a way that gives the best retrieval for a given query.
\end{itemize}

\section{Experiments}
Recall from Sections~\ref{sec:task-defn} and \ref{sec:dense-retrieval-methods} that we had two corpora $D_1$ (seen) and $D_2$ (unseen), and a dataset $Q_1$ of query-relevant passage tuples over $D_1$. We also have test queries for which one passage needs to be retrieved from $D_1$ and $D_2$ each. 

\begin{table*}[tbph]
\centering
\resizebox{0.9\textwidth}{!}{%
\begin{tabular}{|l|rr|rr|rr|}
\hline
 & \multicolumn{2}{c|}{Walmart--Amazon (k=10)} & \multicolumn{2}{c|}{Amazon--Google (k=10)} & \multicolumn{2}{c|}{ConcurrentQA (k=100)} \\ \cline{2-7} 
Known distribution $\rightarrow$ & \multicolumn{1}{c}{Walmart} & \multicolumn{1}{c|}{Amazon} & \multicolumn{1}{c}{Amazon} & \multicolumn{1}{c|}{Google} & \multicolumn{1}{c}{Wikipedia} & \multicolumn{1}{c|}{Enron*} \\ \hline
Na\"ive Merging & 58.71/45.99 & 75.26/62.16 & 67.76/48.87 & 70.26/52.48 & 52.00/09.11 & 80.00/21.23 \\ \hline
Per-task allocation 0.0 & 41.23/33.07 & 45.35/39.05 & 44.05/35.03 & 44.40/35.02 & 51.00/09.10 & 74.00/18.18 \\
Per-task allocation 0.1 & 55.63/45.07 & 69.51/60.42 & 61.25/48.59 & 64.14/51.42 & 55.00/09.15 & 85.00/21.27 \\
Per-task allocation 0.2 & 58.54/45.98 & 76.66/\h{63.09} & 65.09/48.89 & 68.75/52.71 & 54.00/09.16 & 84.00/21.31 \\
Per-task allocation 0.3 & 59.93/46.13 & 78.80/63.01 & 67.11/49.02 & 70.26/52.71 & 54.00/09.17 & 87.00/21.37 \\
Per-task allocation 0.4 & 61.61/\h{46.62} & 80.31/63.04 & 68.19/49.06 & 71.21/\h{52.74} & 55.00/09.20 & 88.00/21.39 \\
Per-task allocation 0.5 & \h{62.02}/46.61 & \h{80.72}/62.97 & \h{68.84}/\h{49.11} & \h{71.47}/52.72 & 56.00/09.22 & \h{88.00}/21.43 \\
Per-task allocation 0.6 & 61.44/46.45 & 80.49/62.84 & 68.62/49.10 & 70.65/52.56 & \h{56.00/09.26} & 87.00/21.49 \\
Per-task allocation 0.7 & 61.15/46.33 & 80.37/62.74 & 67.54/48.95 & 69.22/52.34 & 53.00/09.23 & 87.00/\h{21.58} \\
Per-task allocation 0.8 & 56.10/45.46 & 76.13/62.17 & 59.05/46.58 & 59.40/49.77 & 48.00/09.17 & 83.00/21.63 \\
Per-task allocation 0.9 & 28.11/20.80 & 41.11/35.50 & 30.17/23.80 & 31.08/26.07 & 46.00/09.18 & 80.00/21.92 \\
Per-task allocation 1.0 & 28.11/20.80 & 41.11/35.50 & 30.17/23.80 & 31.08/26.07 & 17.00/01.29 & 53.00/10.88 \\ \hline
Per-query allocation & 67.60/52.80 & 85.13/71.54 & 72.89/58.07 & 74.53/60.26 & 61.00/10.48 & 88.00/30.11 \\ \hline
\end{tabular}

}
\caption{Various allocation strategies evaluated on Walmart--Amazon and \concurrentqa. The reported numbers in each cell are Recall@k and average precision (AP). The chosen k is 10 for Walmart--Amazon, Amazon--Google and 100 for \concurrentqa.}
\label{tab:results}
\end{table*}

\subsection{Training Details}
We use a shared RoBERTa base encoder as the query and passage encoder. We fine-tune it on $Q_1$ for 50 epochs using Adam optimizer with a learning rate of $5e-5$ with warmup for 10\% of the iterations and a batch size of 64 on four TITAN V GPUs. Training take between 4-10 hours depending on the dataset. Further, we truncate queries to 70 tokens and passages to 300. We also present results on fine-tuning \texttt{all-MiniLM-L6-v2} model from \texttt{sentence-transformers} \citep{reimers-2019-sentence-bert} which is a model specifically trained to generate embeddings of sentences and paragraphs for clustering and semantic search.

\subsection{Evaluation}

\begin{figure}[htpb]
    \centering
    \includegraphics[width=0.5\textwidth]{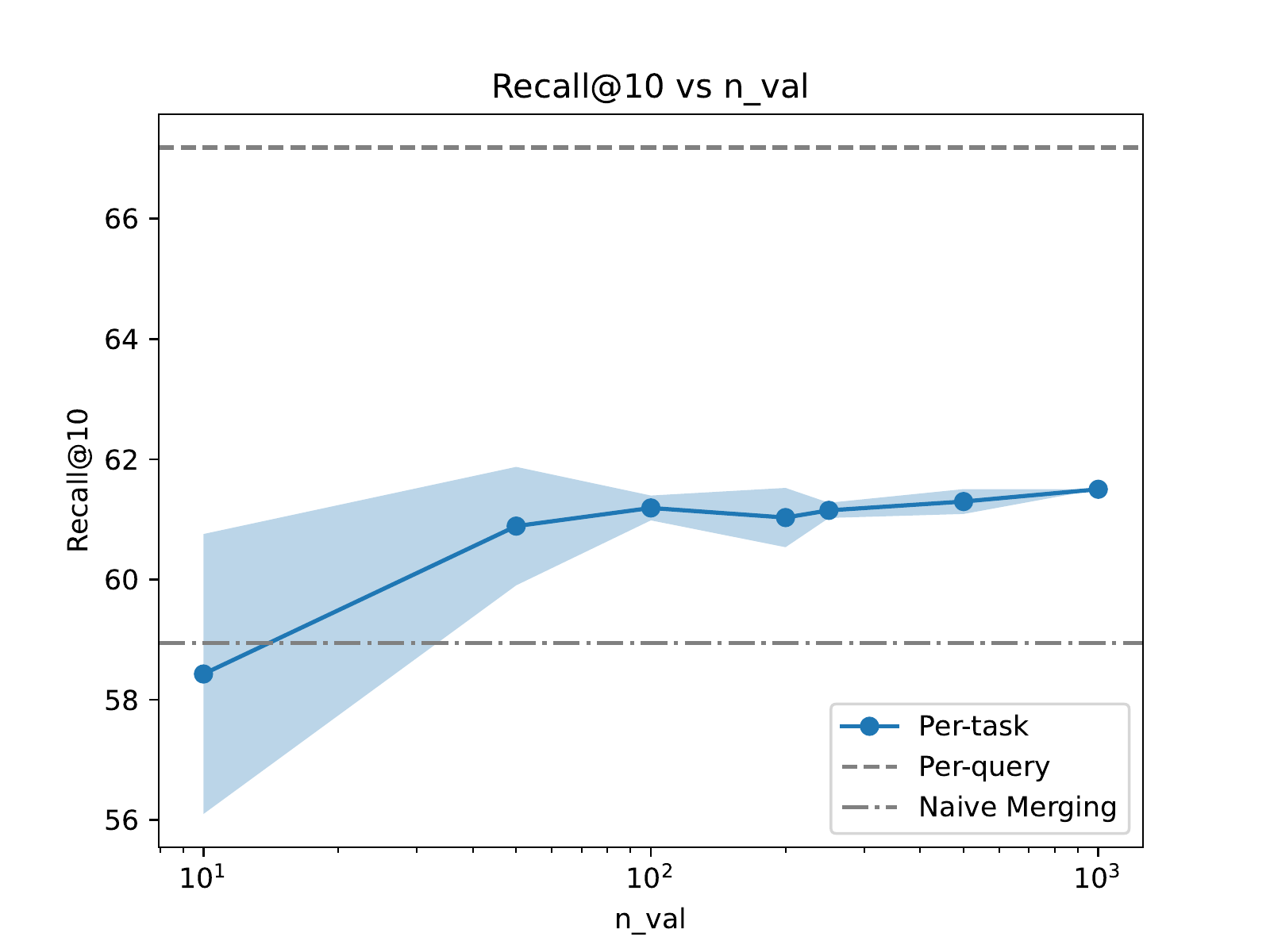}
    \caption{Test accuracy for the best per-task fraction as determined on the validation set of varying size.}
    \label{fig:val}
\end{figure}

Given a query $q$, we compute its embedding $E_Q(q)$, and find the dot product similarity with the embeddings of passages in $D_1$ and $D_2$. We then retrieve some number of passages with the highest similarity scores depending on the allocation strategy to get $k$ candidate passages $\mcal{C}_q$. Finally, the Recall@$k$ is reported. Recall is the fraction of relevant passages retrieved by the model. In our experiments, for a single query, it is $1$ if both passages are retrieved, $0.5$ if one is retrieved and $0$ otherwise. Further, for user facing applications, the rank at which a result is presented is also important. This aspect is not captured by Recall and hence we also report the average precision (AP). Given a list of ranked items, the AP is computed as $\operatorname{AP} = 1/N \sum_{k} P(k) \times rel(k)$ where $N$ is the number of relevant passages for a query (=2 for us), $P(k)$ is the Precision@k and $rel(k)$ is 1 if the passage at rank $k$ is relevant to the query, else 0.

\begin{figure*}[tbp]
    \begin{minipage}[b]{0.56\linewidth}
        \centering
        \includegraphics[width=\linewidth]{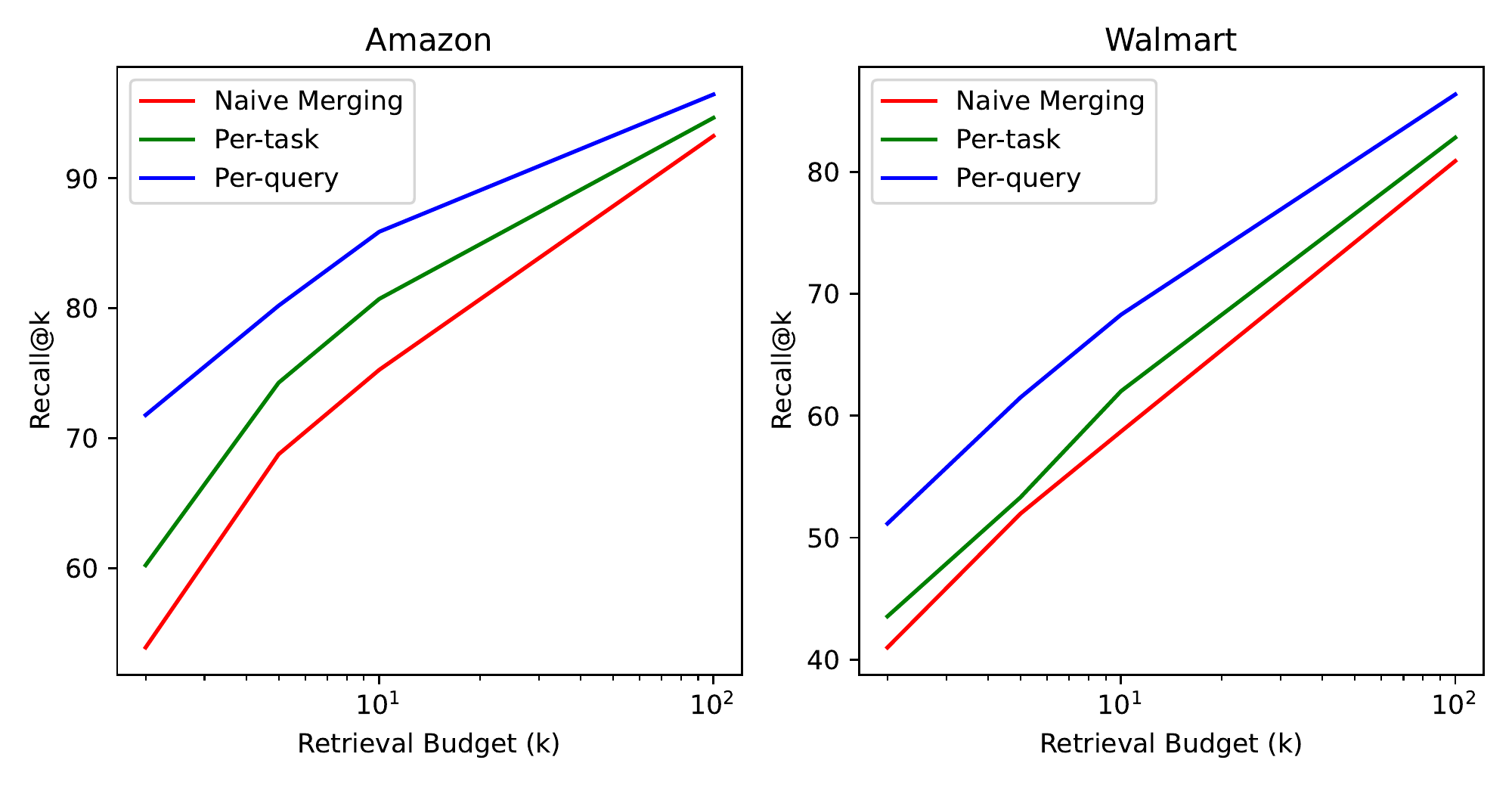}
        \caption{Effect of increasing retrieval budget $k$. With increasing $k$, difference between the three methods decreases while the relative ordering remains the same.}
        \label{fig:effect-k}
    \end{minipage}
    \hfill
    \begin{minipage}[b]{0.41\linewidth}
        \centering
        \includegraphics[width=\linewidth]{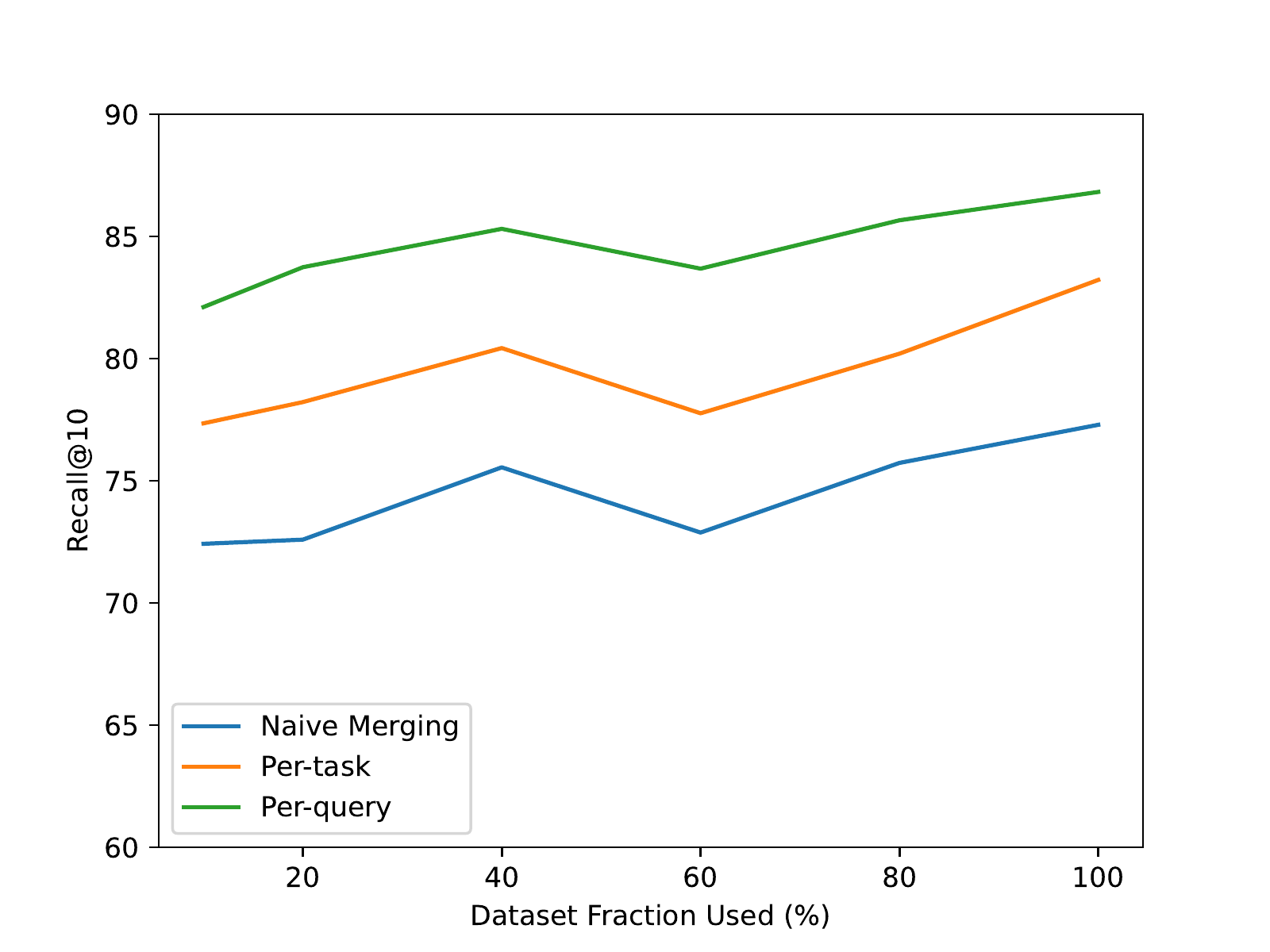}
        \caption{Effect of training data size. Recall increases with training set size but gap between methods remains similar.}
        \label{fig:effect-datasize}
    \end{minipage}
\end{figure*}

\subsection{Results}

\paragraph{Oracle quality} The results on the three datasets is shown in Table~\ref{tab:results}. The Recall@k and average precision values are reported. We choose k=10 for Walmart--Amazon and Amazon--Google datasets, and k=100 for \concurrentqa\ due its much larger corpus size. The effect of changing $k$ is investigated in Section~\ref{sec:analysis}.

We can see that per-task allocation performs better than the na\"ive baseline across all datasets achieving upto 8 points higher recall. As one would expect, performance improves with increasing allocation fraction up to a point before decreasing. Both extremes are not good since the retrieval budget gets allocated to a single corpus. It is interesting to note that the fractions close to 0.5 work best across different settings even though the corpus sizes are imbalanced. Further, on some datasets, the best per-task allocation achieves recall close to that of the per-query allocation while on some there is a gap of 5+ points. We hope that future research on this task will help close this gap.

\begin{figure}[htpb]
    \centering
    \includegraphics[width=0.5\textwidth]{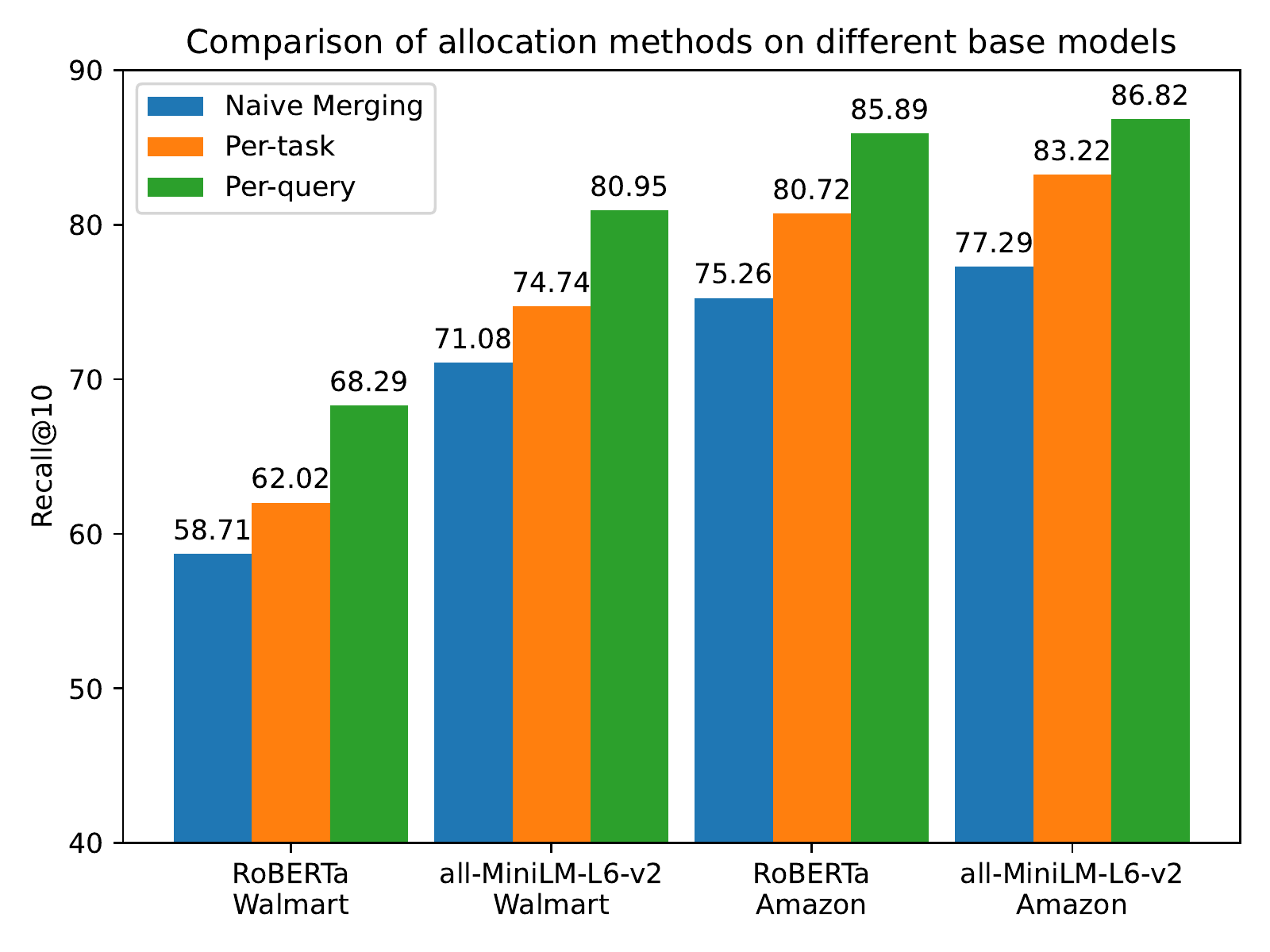}
    \caption{Effect of choice of pretrained model. \texttt{all-MiniLM-L6-v2} performs better than RoBERTa but relative ordering between methods remains the same.}
    \label{fig:effect-models}
\end{figure}

\paragraph{Predicting the Allocation} Next, we consider a simple baseline for \textit{predicting} the allocation fractions across distributions. We propose to determine the fractions from a small seed set of question-passage pairs and report evaluations in Figure~\ref{fig:val}. We sample seed pairs from the full benchmark and report results over pairs that are not included in the seed set. In Figure~\ref{fig:val}, for the Walmart-Amazon dataset with Walmart as the known distribution, we vary the size of the seed set $\in \{10^1,\ldots 10^3\}$ and plot the average result over 10 randomly sampled seed sets per data point. We observe that as few as 10-100 seed pairs are sufficient to select an allocation fraction that outperforms naive merging. Increasing the number of seed pairs corresponds to lower variance in quality across seed sets. As can be seen in Figure~\ref{fig:val-all}, these observations remain consistent over the other datasets as well.

Future work may study unsupervised and increasingly sample-efficient strategies for predicting the allocation fractions. Further, there is a large gap between the predicted allocation fraction and the per-query oracle quality, so it might be important to investigate strategies for predicting the allocation fraction at a per-query granularity.

\section{Analysis}
\label{sec:analysis}

In this section, we evaluate the effect of various hyperparameters and design choices on the performance of our method. First, let us look at the \textbf{effect of increasing the retrieval budget $k$}. As we can see in Figure~\ref{fig:effect-k}, recall increases as we increase $k$ for all the three methods. The per-task allocation numbers plotted here are for the best fraction for that choice of $k$. An interesting observation is that the difference between the three methods decreases as we increase $k$. This is probably due to the fact that when a large budget is available, the correct passage will be retrieved somewhere in the ranked list of passages even though it might have a somewhat lower similarity score.

Next, we investigate the \textbf{effect of training data size} on the performance of the three methods in Figure~\ref{fig:effect-datasize}. We plot the Recall@10 values against increasing fractions of the training set being used for Walmart--Amazon. The model used here was \texttt{all-MiniLM-L6-v2} since it was smaller than RoBERTa and hence faster to train.
As one would expect, the recall increases as more data becomes available but the gains are quite modest. This indicates that even small amount of training data from the known distribution $\mcal{D}_1$ should be enough to fine-tune the encoders.\footnote{Though not related to our task, it might be interesting to evaluate the implications of having a few samples from $\mcal{D}_2$ since it is appears that fine-tuning the encoders does not require a lot of data.}

Finally, we evaluate the \textbf{effect of the choice of pretrained model} by checking if the observations made when fine-tuning RoBERTa also hold for other models. For this, we fine-tune the \texttt{all-MiniLM-L6-v2} model on the Walmart--Amazon dataset and compare the performance of the baseline, the best per-task allocation and per-query allocation in Figure~\ref{fig:effect-models}. Being trained to generate sentence embeddings for semantic search, \texttt{all-MiniLM-L6-v2} performs better than RoBERTa but the relative ordering between the three methods remains the same indicating that our observations can hold across different models.

\section{Conclusion}
In this work, we formalize an underexplored information retrieval task where the retriever applied to several different distributions at inference time, some of which are unseen during training. We created benchmarks for this task and evaluated several simple retrieval methods and models on these benchmarks. We show that these simple methods work well obtaining up to 8 points improvement in Recall@10 over baselines and an average of 3.8 points improvement. 

Limitations of our work include the relatively small size of some of the datasets making fine-tuning and evaluation challenging. Future work can try to create larger datasets particularly suited for this task instead of adapting existing datasets from other tasks. More powerful allocation methods can also be developed. Finally, we had restricted to two distributions. Future work can try extending our study to a larger number of distributions, which we hypothesize will compound the challenge presented by this task considerably.

\bibliographystyle{ACM-Reference-Format}
\bibliography{anthology,custom}

\appendix

\section{Dataset Properties}

\begin{figure*}[!htpb]
    \centering
    \includegraphics[width=0.8\textwidth]{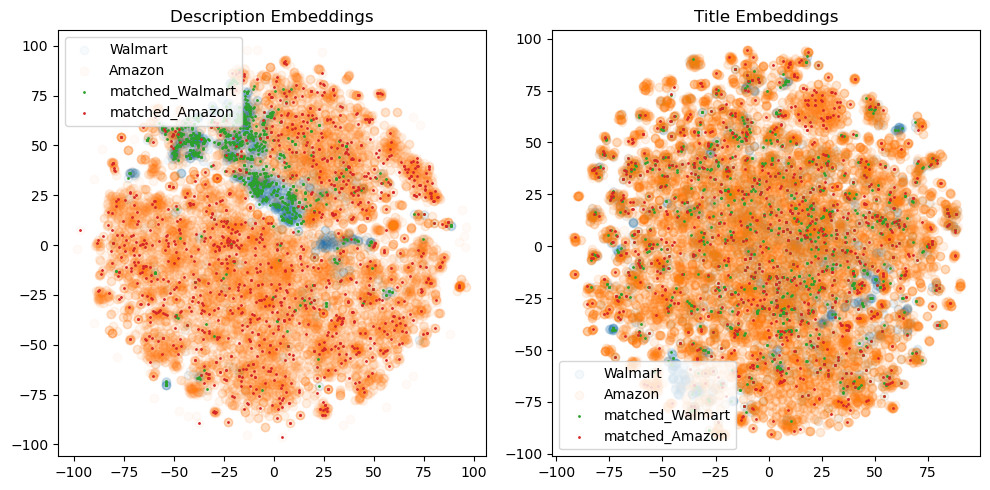}
    \caption{t-SNE plot of the BERT embeddings for the Walmart-Amazon product descriptions and title. It can be seen that the descriptions form two separate distributions while the titles do not.}
    \label{fig:bert-tsne}
\end{figure*}

We would like our datasets to satisfy the two properties to be useful for the multi distribution retrieval task. 
First, we require the presence of two distinct distributions, i.e., the retrieval corpora $D_1$ and $D_2$ should have distinct distributional properties. We plot the t-SNE of BERT embeddings of the passages in the Walmart-Amazon corpus in the Figure~\ref{fig:bert-tsne} to verify this. Further we can see that the queries plotted in the right form a single distribution.
Second, we require that the retrieval from the chosen corpora is non-trivial. We quantify this as retrieval from chosen corpora requiring some in-domain training to achieve reasonable performance. We plot the distribution of zero-shot retrieval ranks using BERT embeddings on the Walmart-Amazon dataset to show this.

\begin{figure*}[!htpb]
    \centering
    \includegraphics[width=0.8\textwidth]{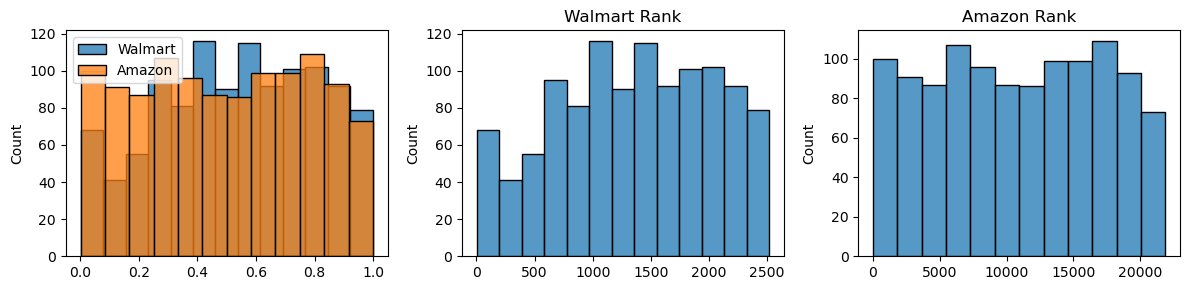}
    \caption{Distributions of ranks of zero-shot retrieval using BERT embeddings. The left figure is the normalized rank while the other two have the actual ranks.}
    \label{fig:bert-ranks}
\end{figure*}

\section{Additional Results on Predicting Allocation Fraction}
\begin{figure*}
    \centering
    \includegraphics[width=\textwidth]{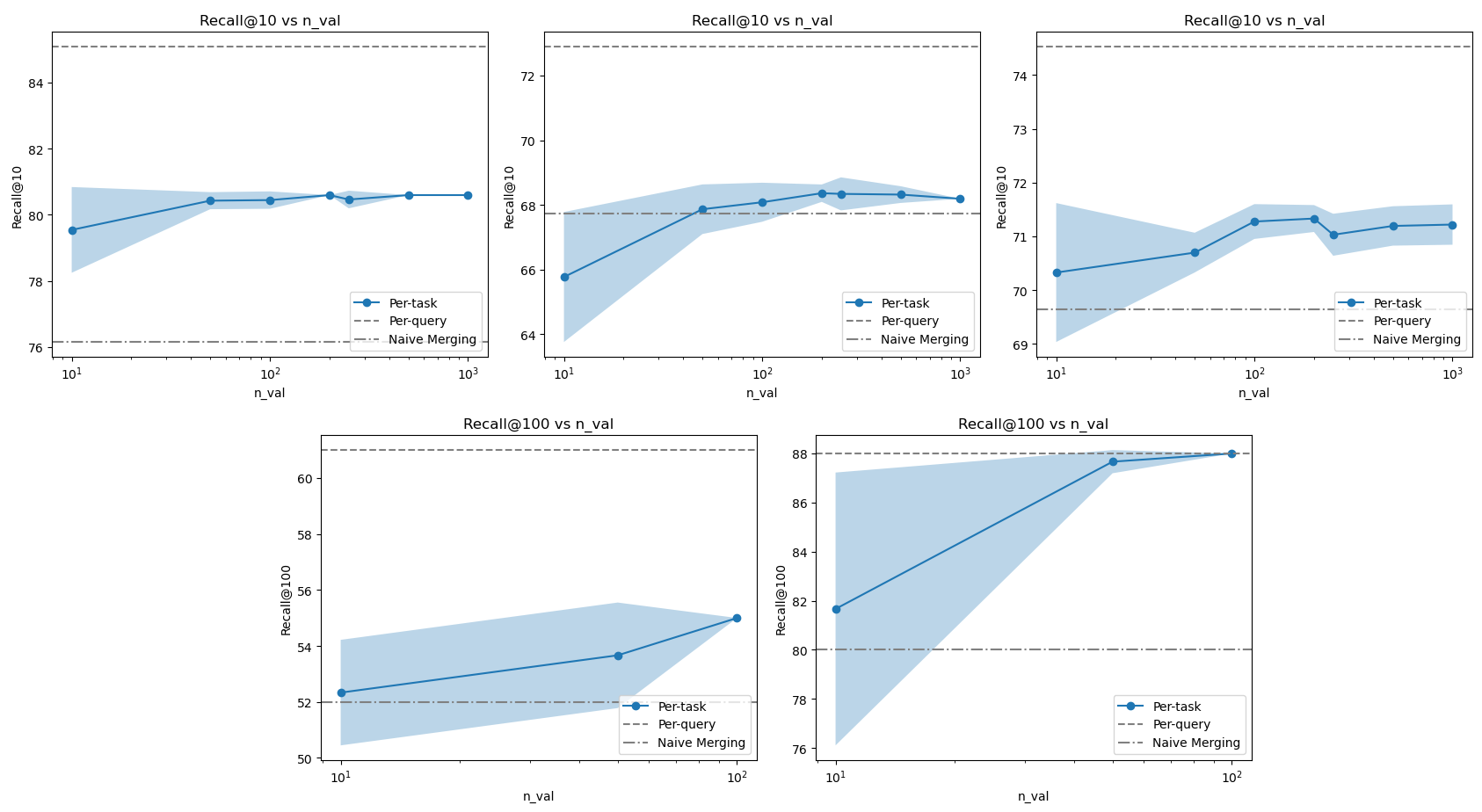}
    \caption{Test accuracy for the best per-task fraction as determined on the validation set of varying size. Left-to-right top-to-bottom are plots for Walmart-Amazon (Amazon), Amazon-Google (Amazon), Amazon-Google (Google), \concurrentqa (Wikipedia) and \concurrentqa (Enron) where the domain in brackets is known.}
    \label{fig:val-all}
\end{figure*}

Figure~\ref{fig:val} had presented the results on predicting the allocation on the Walmart-Amazon dataset with Walmart as the known distribution. We provide the plots for the remaining datasets and settings in Figure~\ref{fig:val-all} to show that the observations remain consistent over datasets and for completeness.

\end{document}